\documentstyle[aps,epsfig]{revtex}
\begin{document}

\title{Algebraic Solutions of an Extended Pairing Model for
Well-Deformed Nuclei}
\author{Feng Pan$^{1,2}$, V. G. Gueorguiev\footnote{
On leave of absence from Institute of Nuclear Research and Nuclear Energy, Bulgarian Academy of Sciences, Sofia 1784, Bulgaria.}$^{2}$, and J. P. Draayer$^{2}$}
\address{ $^{1}$Department of Physics, Liaoning Normal University,
Dalian, 116029, P. R. China\\
$^{2}$Department of Physics and Astronomy, Louisiana State University, Baton
Rouge, LA 70803}
\maketitle

\vspace{0.5cm}
\begin{abstract}
A Nilsson mean-field plus extended pairing interaction Hamiltonian with
many-pair interaction terms is proposed.  Eigenvalues of the extended
pairing model
are easy to obtain. Our investigation shows that one- and two-body interactions
continue to dominate the dynamics for relatively small values of the
pairing strength.  As the strength of the pairing interaction grows, however,
the three- and higher many-body interaction terms grow in importance.
   A numerical study of even-odd mass differences
in the $^{154-171}$Yb isotopes shows that the extended pairing model is
applicable to well-deformed nuclei.

\end{abstract}



Pairing is an important residual interaction that is used in nuclear
physics as well as in other fields such as the study of metallic clusters.
The Bardeen-Cooper-Schrieffer (BCS) \cite{BCS} and Hartree-Fock-Bogolyubov
(HFB) \cite{HFB} methods for finding approximate solutions are well known.
The limitations of these methods when applied in nuclear physics are also
well understood. First of all, not only is the number of nucleons in a
nucleus typically small, the number of valence particles ($n\sim 10$) which
dominate the behavior of low-lying states is too few to support the
underlying assumptions of the approximations, that is, particle
number fluctuations are non-negligible. As a result, particle
number-nonconservation effects can lead to serious problems such as
spurious states, nonorthogonal solutions, etc. Furthermore, an
essential feature of pairing correlations are differences between
neighboring even and odd mass nuclei, which are driven mainly by Pauli
blocking effects, and it is difficult to treat even-odd differences with
either the BCS or HFB theories because different quasi-particle bases
must be introduced for different blocked levels. Another problem with
approximate treatments of the pairing Hamiltonian is related to the fact
that both the BCS and the HFB approximations break down for an important
class of physical situations. A remedy that uses particle number
projection techniques complicates the algorithms considerably and does
not help to achieve a better description of the higher-excited part of
the spectrum of the pairing Hamiltonian. Similar conclusions were found
to hold within the context of the energy spectra of nano-scale metallic
grains \cite{Black CT,Ralph DC}.

Driven by the importance of having exact solutions of the pairing
Hamiltonian, much attention and progress, building on Richardson's early
work \cite{Richarson'63a,Richarson'63b,Richarson'64} and extensions to it
based on the Bethe ansatz, has been made in the past few years
\cite{Feng'98a,Feng'98b,Feng'99,Feng'00a,Dukelsky'01,Dukelsky'02,Zhou'02}.
For all these algebraic Bethe ansatz approaches, the solutions are
provided by a set of highly non-linear Bethe Ansatz Equations (BAEs).
Though these applications demonstrate that the pairing problem is exactly
solvable, solutions of these BAEs are not easy and normally require
extensive numerical work, especially when the number of levels and valence
pairs are large.  This limits the applicability of the methodology to
relatively small systems; it cannot be applied to large systems such as
well-deformed nuclei.


The standard pairing Hamiltonian for well-deformed nuclei is given
by
\begin{equation}
\hat{H}=\sum_{j=1}^{p}\epsilon _{j}n_{j}-G\sum_{i,j=1}^{p}a_{i}^{+}a_{j},
\label{eqno1}
\end{equation}
where $p$ is the total number of Nilsson levels considered, $G>0$ is the
overall pairing strength, $\epsilon _{j}$ is single-particle energies taken
from the Nilsson model, $n_{j}=c_{j\uparrow }^{\dagger }c_{j\uparrow
}+c_{j\downarrow }^{\dagger }c_{j\downarrow }$ is the fermion number
operator for the $j$-th Nilsson level, and $a_{i}^{+}=c_{i\uparrow
}^{\dagger }c_{i\downarrow }^{\dagger }$ ( $a_{i}=(a_{i}^{+})^{\dagger
}=c_{i\downarrow }c_{i\uparrow }$) are pair creation (annihilation)
operators. The up and down arrows in these expressions refer to time-reversed
states. Since each Nilsson level can only be occupied by one pair due to the
Pauli Principle, the Hamiltonian (\ref{eqno1}) is also equivalent to a finite
site hard-core Bose-Hubbard model with infinite range one-pair hopping
and infinite on-site repulsion. Specifically, the operators
$a_{i}^{+}$, $a_{i}$, and $n_{i}^{a}=n_{i}/2$ satisfy the following
hard-core boson algebra:

\begin{equation}
(a_{i}^{+})^{2}=0,~~[a_{i},a_{j}^{+}]=\delta
_{ij}(1-2n_{i}^{a}),~~[a_{i}^{+},a_{j}^{+}]=[a_{i},a_{j}]=0.  \label{eqno2}
\end{equation}

As an extension of (\ref{eqno1}), we construct the following new
(extended) pairing Hamiltonian:

\begin{eqnarray}
\hat{H}&=&\sum_{j=1}^{p}\epsilon _{j}n_{j}-
G\sum_{i,j=1}^{p}a_{i}^{+}a_{j}
\label{eqno3} -G\left( \sum_{\mu =2}^{\infty }{\frac{1}{{{(\mu !)}^{2}}}}
\sum_{i_{1}\neq i_{2}\neq \cdots \neq i_{2\mu
}}a_{i_{1}}^{+}a_{i_{2}}^{+}\cdots a_{i_{\mu }}^{+}a_{i_{\mu +1}}a_{i_{\mu
+2}}\cdots a_{i_{2\mu }}\right).
\end{eqnarray}
Besides the usual Nilsson mean-field and the standard pairing interaction
(\ref{eqno1}), this form includes many-pair hopping terms that allow
nucleon pairs to simultaneously scatter (hop) between and among different
Nilsson levels. With this extension in place, we will show that the model
is exactly solvable.

Because of the infinite on-site repulsion, the infinite sum in
(\ref{eqno3}) truncates for $\mu \leq p.$ It is also easy to see that each
term of the form $a_{i}^{+}\cdots a_{j}^{+}$ that forms eigenstates of
(\ref{eqno3}) should enter with different indices $i\neq \cdots \neq j$.
Let $|j_{1},\cdots ,j_{m}\rangle $ be the pairing vacuum state that
satisfies

\begin{equation}
a_{i}|j_{1},\cdots ,j_{m}\rangle =0  \label{eqno4}
\end{equation}
for $1\leq i\leq p$, where $j_{1},j_{2},\cdots ,j_{m}$ indicates those
$m $ levels that are occupied by single nucleons. Any singly-occupied
state is blocked by the Pauli principle.

Following the algebraic ansatz used in \cite{Feng'00b}, one can write
$k$-pair eigenstates of (\ref{eqno3}) as

\begin{equation}
|k;\zeta ;j_{1},\cdots ,j_{m}\rangle =\sum_{1\leq i_{1}<i_{2}<\cdots
<i_{k}\leq p}C_{i_{1}i_{2}\cdots i_{k}}^{(\zeta
)}a_{i_{1}}^{+}a_{i_{2}}^{+}\cdots a_{i_{k}}^{+}|j_{1},\cdots ,j_{m}\rangle ,
\label{eqno5}
\end{equation}
where $C_{i_{1}i_{2}\cdots i_{k}}^{(\zeta )}$ is an expansion
coefficient that needs to be determined, and the strict ordering to the
indices $i_{1},i_{2},\cdots ,i_{k}$ reminds us that double
occupation is not allowed. It is always assumed that the level indices
$j_{1},~j_{2},\cdots ,j_{m}$ should be excluded from the summation in
(\ref{eqno5}). Since the formalism for even-odd systems is similar, in
the following, we focus on the even-even seniority zero case.

The expansion coefficient $C_{i_{1}i_{2}\cdots i_{k}}^{(\zeta )}$ can
be expressed very simply as

\begin{equation}
C_{i_{1}i_{2}\cdots i_{k}}^{(\zeta )}=\frac{1}{{1-x^{(\zeta )}\sum_{\mu
=1}^{k}\epsilon _{i_{\mu }}}},  \label{eqno6}
\end{equation}
where, similar to the results given in the Bethe ansatz approach,
$x^{(\zeta )}$ is a c-number that is to be determined. To prove that the
algebraic ansatz given in (\ref{eqno5}) and (\ref{eqno6}) are consistent,
one may directly apply Hamiltonian (\ref{eqno3}) on the $k$-pair state
(\ref{eqno5}). Using the hard-core boson algebraic relation given by
(\ref{eqno2}) and a procedure that is similar to that used in Ref.
\cite{Feng'99} for finding exact solution of a Heisenberg algebra
Hamiltonian, one can determine rather easily that for the mean-field part of
the Hamiltonian (\ref{eqno3})

\begin{equation}
\sum_{j}\epsilon _{j}n_{j}|k;\zeta ;0\rangle =\frac{2}{{x^{(\zeta )}}}
\left( |k;\zeta ;0\rangle -\sum_{1\leq i_{1}<i_{2}<\cdots <i_{k}\leq
p}a_{i_{1}}^{+}a_{i_{2}}^{+}\cdots a_{i_{k}}^{+}|0\rangle \right) ,
\label{eqno7}
\end{equation}
and for the extended pairing part of the Hamiltonian (\ref{eqno3})

\[
\left( \sum_{i}a_{i}^{+}a_{i}+\sum_{\mu =1}^{\infty }{\frac{1}{{{(\mu !)}^{2}
}}}\sum_{i_{1}\neq i_{2}\neq \cdots \neq i_{2\mu
}}a_{i_{1}}^{+}a_{i_{2}}^{+}\cdots a_{i_{\mu }}^{+}a_{i_{\mu +1}}a_{i_{\mu
+2}}\cdots a_{i_{2\mu }}\right) |k;\zeta ;0\rangle =
\]
\begin{equation}
\left( \sum_{1\leq i_{1}<i_{2}<\cdots <i_{k}\leq p}C_{i_{1}i_{2}\cdots
i_{k}}^{(\zeta )}\right) \sum_{1\leq i_{1}<i_{2}<\cdots <i_{k}\leq
p}a_{i_{1}}^{+}a_{i_{2}}^{+}\cdots a_{i_{k}}^{+}|0\rangle +(k-1)|k;\zeta
;0\rangle  \label{eqno8}
\end{equation}
By combining Eqs. (\ref{eqno7}) and (\ref{eqno8}), the $k$-pair excitation
energies of (\ref{eqno3}) are given by:

\begin{equation}
E_{k}^{(\zeta )}=\frac{2}{x{^{(\zeta )}}}-G(k-1),  \label{eqno9}
\end{equation}
where the undetermined variable $x^{(\zeta )}$ is given by
\begin{equation}
{\frac{2}{{x^{(\zeta )}}}}+\sum_{1\leq i_{1}<i_{2}<\cdots <i_{k}\leq p}
{\frac{G}{(1-{x^{(\zeta )}\sum_{\mu =1}^{k}\epsilon _{i_{\mu }})}}=0}.
\label{eqno10}
\end{equation}
The additional quantum number $\zeta $ now can be understood as the
$\zeta$-th solution of (\ref{eqno10}). Similar results for even-odd systems can
also be derived by using this approach except that the index $j$ of the
level occupied by the single nucleon should be excluded from the summation
in (\ref{eqno5}) and the single-particle energy term $\epsilon _{j}$
contributing to the eigenenergy from the first term of (\ref{eqno3}) should
be included. Extensions to many broken-pair cases are straightforward.

Comparing Eqs. (\ref{eqno9}) and (\ref{eqno10}) to exact solutions of the
Heisenberg algebraic Hamiltonian with a one-body interaction
\cite{Feng'99}, one can regard the operator product
$a_{i_{1}}^{+}a_{i_{2}}^{+}\cdots a_{i_{k}}^{+}$ in (\ref{eqno5}) as a
`grand' boson. The corresponding `single-particle energy' of the `grand'
boson is $E_{i_{1}i_{2}...i_{k}}=\sum_{\mu =1}^{k}2 {\epsilon }_{i_{\mu
}}$, since (\ref{eqno10}) and the eigenstates (\ref{eqno5}) are similar
to those for a multi-boson system with a one-body interaction as shown in
\cite{Feng'99}, even though the Hamiltonians are totally different.
Although the eigenstates (\ref{eqno5}) are not normalized, but they can
be normalized easily by using a standard procedure. The eigenstates
(\ref{eqno5}) with different roots given by (\ref{eqno10}) are mutually
orthogonal, a result that can be proven as in \cite{Feng'99}.

Eigenenergies of the standard pairing model are expressed in terms of $k$
variables that satisfy $k$ coupled nonlinear equations that are difficult to
solve numerically, especially when the number of pairs $k$ and number of
levels $p$ are large. In contrast to BAE solutions to the standard pairing
model, there is but a single variable $x^{(\zeta )}$ in the extended model.
It should be noted that the solution (\ref{eqno10}) is only valid when all
combinations of the single-particle energies
$\sum_{\mu =1}^{k}\epsilon _{i_{\mu }}$ are different for $k$-pair
excitation cases. Fortunately, this is always the case when single-particle
energies are generated from the Nilsson model. In that case, (\ref{eqno10})
should have $\frac{p!}{(p-k)!p!}$ distinct roots, which could be a very
large number for an entire deformed major shell.

If (\ref{eqno10}) is
rewritten in terms of $z^{(\zeta )}=2/(G x^{(\zeta )})$ and the
dimensionless energy of the `grand' boson
$\tilde{E}_{i_{1}i_{2}...i_{k}}=\sum_{\mu =1}^{k}\frac{2 {\epsilon
}_{i_{\mu }}}{G},$ (\ref{eqno10}) takes the form:
\begin{equation}
1=\sum_{1\leq i_{1}<i_{2}<\cdots <i_{k}\leq p}{\frac{1}{(
\tilde{E}_{i_{1}i_{2}...i_{k}}{-}z^{(\zeta )}{)}}}.
\label{eqno11}
\end{equation}
Since there is only one variable $z^{(\zeta )}$ in (\ref{eqno11}), the
zero points of the function can easily be shown graphically, which is very
similar to the one-pair solution appearing in the TDA and RPA
approximations with separable potentials\cite{HFB}. From expression
(\ref{eqno11}), it is clear that any solution $z^{(\zeta )}$ of
(\ref{eqno11}) is located between two nearby values of the dimensionless
`grand' boson energy
$\tilde{E}_{i_{1}i_{2}...i_{k}}=\sum_{\mu =1}^{k}\frac{2{\epsilon
}_{i_{\mu }}}{G}$
and the smallest solution $z^{(\zeta )}$ would be smaller than
$\min(\{\tilde{E}_{i_{1}i_{2}...i_{k}}\}).$

\begin{figure}[tbp]
\centerline{\hbox{\epsfig{file=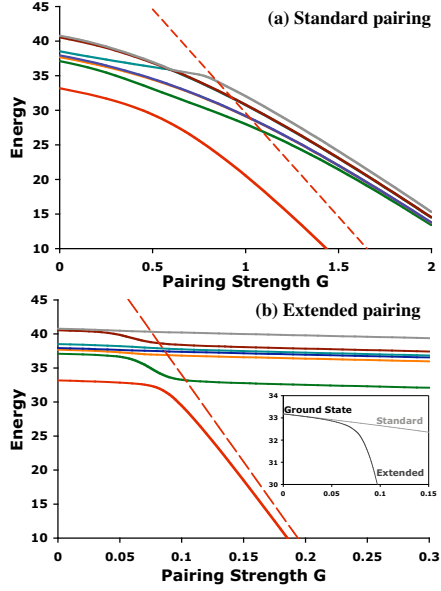, width=6.5cm}}}
\caption{ (a) Spectral structure of the standard pairing interaction
given by Eq.(\ref{eqno1}), and (b) spectral structure of the extended
pairing interaction given by Eq.(\ref{eqno3}), as functions of the pairing
interaction strength $G$ for
$k=5$ pairs for a system with $p=10$ levels and single-particle energies
$\epsilon _{1}=1.179$, $\epsilon_{2}=2.65$,
$\epsilon_{3}=3.162$, $\epsilon_{4}=4.588$,
$\epsilon_{5}=5.006$, $\epsilon_{6}=6.969$,
$\epsilon_{7}=7.262$, $\epsilon_{8}=8.687$,
$\epsilon_{9}=9.899$, $\epsilon_{10}=10.201$,
where the single-particle energies and $G$ are given in arbitrary units.
The straight dash line is the expectation value of the Hamiltonian in the
pure pairing ($\epsilon_{i}=0$) ground state.}
\label{PairingSpectra}
\end{figure}

It is important to understand the differences between the
extended pairing introduced by
(\ref{eqno3}) and the standard pairing given in (\ref{eqno1}). For this
purpose, let us consider a simple example in which there are $p=10$
levels and the single-particle energies are given by
$\epsilon _{i}=i+\chi _{i}$ for $i=1,2,\cdots ,10$, where $\chi _{i}$ are
random numbers within the interval $(0,1)$ to avoid accidental degeneracy
required for exact solvability, and the pairing strength $G$ is
allowed to vary from $0.01$ to $0.10$.
Fig. \ref{PairingSpectra} shows the lowest few energies of the standard
and extended pairing models. From this graph it is very clear that there
are essential differences in the spectral structure of these two models.
As shown by Inset (b), the extended pairing model very rapidly develops a
paired ground-state configuration which is strongly dependent on the
pairing strength $G$. In this case the transition from mean-field
eigenstates to pairing eigenstates is very sharp and fast, while the
standard pairing model, Inset (a), exhibits a much slower and smoother
transition. The quantitative difference in the two spectra,  with the
extended pairing case showing a much stronger dependence on $G$ than for
standard pairing, is a very clear distinguishing characteristic and can
be used to explore cases where the extended pairing concept might be more
relevant and appropriate than the standard pairing model.

\begin{figure}[tbp]
\centerline{\hbox{\epsfig{file=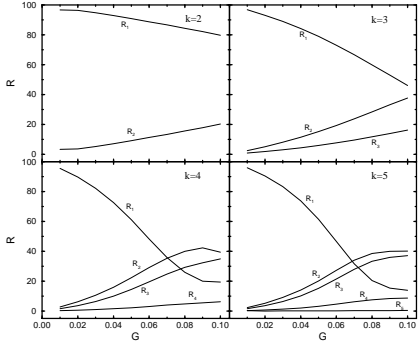, width=6.5cm}}}
\caption{ Ratios $R_{\mu }$($\%$) with $\mu =1,2,\cdots ,5$ as function of
the pairing interaction strength $G$ for $k=2,\cdots ,5$ for a system
with $p=10$ levels. The parameters that were used are the same as those
shown in Fig.~\ref{PairingSpectra}.}
\label{Vk Ratios}
\end{figure}

Since there are higher order terms involved in (\ref{eqno3}), it is
important to know whether the dynamics is still dominated by the one-
and two-body interactions, and, if not, under what conditions the higher
order terms can be treated perturbatively. To explore this, we calculated
as a function of $G$ the expectation value
of each higher order term $\langle V_{\mu }\rangle $ defined by:

\begin{equation}
V_{1}=\sum_{i,j}a_{i}^{+}a_{j},~~V_{\mu }={\frac{1}{{(\mu !)^{2}}}}
\sum_{i_{1}\neq i_{2}\neq \cdots \neq i_{2\mu
}}a_{i_{1}}^{+}a_{i_{2}}^{+}\cdots a_{i_{\mu }}^{+}a_{i_{\mu +1}}a_{i_{\mu
+2}}\cdots a_{i_{2\mu }}  \label{eqno12}
\end{equation}
with $\mu =2,3,\cdots $, for $k$-pair ground states. Then, we calculate the
ratio $R_{\mu }=\langle V_{\mu }\rangle /\langle V_{{\rm total}}\rangle $,
where $\langle V_{{\rm total}}\rangle $ is the sum of all terms given in
(\ref{eqno12}). The results up to the half-filled case are shown in Fig.
\ref{Vk Ratios}. It can be seen that the two-body pairing interaction
($V_{1}$) dominates the dynamics of the system as long as the
interaction strength $G$ is small. With increasing interaction strength,
the system is driven mainly by $V_{2},$ less by $V_{3}$ (but it may be
comparable with $V_{2}$), and much less by the higher order terms. As one
would expect, increasing the number of pairs $k$ drives the critical
point where $V_{2}$ becomes dominant towards smaller $G$ values. The
situation and the graphs beyond the half-filled case are qualitatively
similar because of the particle-hole symmetry. The critical point where
$V_{2}$ becomes dominant is actually at higher $G$ values for $p-k$
pairs than for the $k<p/2$ pairs. Even though, higher order terms appear
beyond the half-filled case, these terms are always less important
than $V_{1}$ for small $G$ and $V_{2}$ for big values of $G$.
It is important to note that for large values of $G$ when the dynamics
is dominated by the pairing interaction, and thus independent of $G$,
the $V_{2}$ term dominates followed in importance by the $V_{3}$ term.

\begin{figure}[tbp]
\centerline{\hbox{\epsfig{file=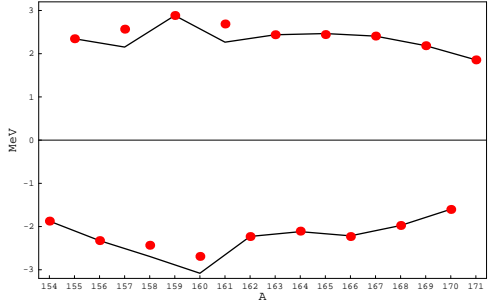, width=6.5cm}}}
\caption{Even-odd mass difference $P(A)=E(A)+E(A-2)-2E(A-1)$ for $^{154-171}
$Yb, where $E(A)$ is the total binding energy, and the dots correspond to
the experimental quantities. The theoretical values for even-odd mass P(A)
are connected by the lines.}
\label{even-odd}
\end{figure}

As an example of an application of the theory to well-deformed nuclei,
we fit even-odd mass differences of the $^{154-171}$Yb isotopes. The
single-particle energies of each nucleus were calculated using the
Nilsson deformed shell model with experimentally evaluated deformation
parameters\cite{Nix JR}. Fig. \ref{even-odd} shows the fitted results in
comparison with the corresponding experimental values\cite{Audi G}.
Except for small deviations for $^{157-161}$Yb, the results are well
reproduced.

The corresponding $G$ values are given in Table \ref{Table1}, from which
one can see that the pairing interaction strength
decreases with increasing number of pairs $k$ from $245$keV for $1$ pair
to $0.0948$keV for $9$ pairs, while the single-particle energy gaps are
always about a few hundreds keV. This situation is characteristic of the
extended pairing model. Overall, the results suggest that the model may
be applicable to well-deformed nuclei. The model may also
be useful in studying pairing phenomena in metallic clusters of nano-scale
size.

\vspace{0.5cm}
\begin{table}[tbph]
\caption{Pairing interaction strength $\vert G\vert$ (keV) used in Fig.
\ref{even-odd} for the $^{154-171}$Yb isotopes.}
\vskip 0.25cm
\begin{tabular}{cccccccccc}
&  &  &  &  &  &  &  &  &  \\
& $^{154}$Yb & $^{156}$Yb & $^{158}$Yb & $^{160}$Yb & $^{162}$Yb & $^{164}$Yb
& $^{166}$Yb & $^{168}$Yb & $^{170}$Yb \\
&  &  &  &  &  &  &  &  &  \\ \hline
$k$ & 1 & 2 & 3 & 4 & 5 & 6 & 7 & 8 & 9 \\ \hline
$G$ & 245 & 41.1 & 6.5 & 3.02 & 1.11145 & 0.4675 & 0.2337 & 0.1376
& 0.094816 \\
\hline\hline
&  &  &  &  &  &  &  &  &  \\
& $^{155}$Yb & $^{157}$Yb & $^{159}$Yb & $^{161}$Yb & $^{163}$Yb & $^{165}$Yb
& $^{167}$Yb & $^{169}$Yb & $^{171}$Yb \\
&  &  &  &  &  &  &  &  &  \\ \hline
$k$ & 1 & 2 & 3 & 4 & 5 & 6 & 7 & 8 & 9 \\ \hline
$G$ & 270.7 &42.0 & 10.9 & 1.0 & 1.471 & 0.649 & 0.3477 & 0.2185 &
0.16113 \\
\end{tabular}
\label{Table1}
\end{table}

\vskip .5cm
We acknowledge support provided by the U.S. National Science Foundation
(0140300), the Natural Science Foundation of China (10175031), and
the Education Department of Liaoning Province (202122024).


\begin{thebibliography}{}
\bibitem{BCS} J. Bardeen, L. N. Cooper, and J. R. Schrieffer, Phys. Rev.
{\bf 108}, 1175 (1957).

\bibitem{HFB}  P. Ring and P. Schuck, {\it The Nuclear Many-Body Problem}
(Springer Verleg, 1980, Berlin).

\bibitem{Black CT}  C. T. Black, D. C. Ralph, and M. Tinkham, Phys. Rev.
Lett. {\bf 76}, 688 (1996).

\bibitem{Ralph DC}  D. C. Ralph, C. T. Black, and M. Tinkham, Phys. Rev.
Lett. {bf 78}, 4087(1996).

\bibitem{Richarson'63a}  R. W. Richardson, Phys. Lett. {\bf 3}, 277 (1963).

\bibitem{Richarson'63b}  R. W. Richardson, Phys. Lett. {\bf 5}, 82 (1963).

\bibitem{Richarson'64}  R. W. Richardson and N. Sherman, Nucl. Phys.
{\bf 52}, 221 (1964).

\bibitem{Feng'98a}  Feng Pan, J. P. Draayer, and W. E. Ormand, Phys. Lett.
{\bf B422}, 1 (1998).

\bibitem{Feng'98b}  Feng Pan and J. P. Draayer, Phys. Lett. {\bf B442}, 7
(1998).

\bibitem{Feng'99}  Feng Pan and J. P. Draayer, Ann. Phys. (NY) {\bf 271},
120 (1999).

\bibitem{Feng'00a}  Feng Pan, J. P. Draayer, and Lu Guo, J. Phys. A: Math.
Gen. {\bf 33}, 1597 (2000).

\bibitem{Dukelsky'01}  J. Dukelsky, C. Esebbag, and P. Schuck, Phys. Rev.
Lett. {\bf 87}, 066403 (2001).

\bibitem{Dukelsky'02}  J. Dukelsky, C. Esebbag, and S. Pittel, Phys. Rev.
Lett. {\bf 88}, 062501 (2002).

\bibitem{Zhou'02}  H. -Q. Zhou, J. Links, R. H. McKenzie, and M. D. Gould,
Phys. Rev. B {\bf 65}, 060502(R) (2002).

\bibitem{Feng'00b}  Feng Pan and J. P. Draayer, J. Phys. A: Math. Gen.
{\bf 33}, 9095 (2000).

\bibitem{Nix JR}  J. R. Nix and K. L. Kratz, Atomic Data Nucl. Data Tables
{\bf 66}, 131 (1997).

\bibitem{Audi G}  G. Audi et al., Nucl. Phys. {\bf A624}, 1 (1997).

\end{thebibliography}
\end{document}